\documentclass[11pt]{article}
\usepackage[margin=1in]{geometry}
\usepackage{graphicx}
\usepackage{booktabs}
\usepackage{amsmath}
\usepackage{amssymb}
\usepackage{cmap}                       
\usepackage[T1]{fontenc}
\usepackage{helvet}                     
\input{glyphtounicode}\pdfgentounicode=1 
\usepackage[expansion=false,protrusion=false]{microtype} 
\DisableLigatures[f]{encoding = T1, family = *} 
\usepackage{natbib}
\usepackage{xcolor}
\definecolor{linkblue}{HTML}{2A5DB0}
\definecolor{boxgray}{HTML}{F0F2F5}
\usepackage[colorlinks=true,citecolor=linkblue,linkcolor=linkblue,urlcolor=linkblue]{hyperref}

\makeatletter
\renewcommand\section{\@startsection{section}{1}{\z@}%
  {-3.5ex \@plus -1ex \@minus -.2ex}{2.0ex \@plus.2ex}%
  {\normalfont\large\sffamily\bfseries}}
\renewcommand\subsection{\@startsection{subsection}{2}{\z@}%
  {-2.5ex\@plus -1ex \@minus -.2ex}{1.2ex \@plus .2ex}%
  {\normalfont\normalsize\sffamily\bfseries}}
\renewcommand\paragraph{\@startsection{paragraph}{4}{\z@}%
  {1.6ex \@plus1ex \@minus.2ex}{-0.6em}%
  {\normalfont\normalsize\sffamily\bfseries}}
\makeatother
\newcommand{\abstractbox}[1]{%
  \begingroup\setlength{\fboxsep}{10pt}%
  \noindent\colorbox{boxgray}{\begin{minipage}{\dimexpr\textwidth-2\fboxsep\relax}#1\end{minipage}}%
  \endgroup}

\begin{document}

\begin{flushleft}
{\sffamily\bfseries\fontsize{20}{24}\selectfont
A global predicted-fMRI drive signal from TRIBE does not predict YouTube replay heatmaps\par}
\vspace{10pt}
\begin{tabular}[t]{@{}l@{\hskip 3.5em}l@{}}
{\sffamily\bfseries Barada Sahu} & {\sffamily\bfseries Shivesh Pandey}\\[1pt]
Cabal AI & Para AI
\end{tabular}
\end{flushleft}

\abstractbox{%
Deep multimodal brain-encoding models now predict fMRI responses to naturalistic video with high
accuracy. Whether their \emph{predicted} neural signals also forecast behavioral engagement is
unknown. We run TRIBE, the winning model of the 2025 Algonauts brain-encoding challenge
(Llama-3.2 + V-JEPA 2 + Wav2Vec-BERT), on 48 YouTube videos and reduce its predicted cortical
response to a per-second engagement curve, the global field power. Correlated against each video's
``most replayed'' heatmap, a passively-collected proxy for which moments viewers return to, the
curve shows no evidence of predicting re-watch behavior. The pooled position-controlled partial
correlation is $+0.058$ (95\% CI $[-0.04, 0.15]$; one-sample $t(47)=1.21$, $p=0.23$),
indistinguishable from zero and not significantly above simple loudness and motion baselines
(loudness $+0.04$, paired $p=0.74$). The raw correlation is also near zero; the moderate values reported for music
videos are a genre-specific intro/onset-replay artifact and do not generalize to other content. The null holds across six cortical-network readouts, signed value/salience ROIs, and
an autocorrelation-preserving permutation test; a supervised leave-one-video-out probe on the
predicted cortex appears to reach $r=0.47$ but this collapses to a shared temporal-shape artifact
under a proper position control. Running the same probe on TRIBE's three input streams reveals at
most a small, borderline content-specific signal in the visual stream (matched vs.\ mismatched
$p\approx0.004$--$0.06$ across feature extractions) and none in audio, text, or the predicted cortex,
tentatively placing what little signal exists in the visual input, upstream of the encoding. The
inter-subject-correlation (ISC) readout, the closest prior positive result at this grain, is
unavailable from the subject-averaged released model, so we fit our own per-subject encoders on
the Algonauts fMRI (validated in-domain at $r\approx0.15$ and cross-domain, Friends$\to$film, at
$r\approx0.10$); the predicted ISC still does not track re-watch ($r=-0.04$, $p=0.34$). We
bound the claim rather than merely fail to reject it: a Bayes
factor gives moderate evidence for the null ($\mathrm{BF}_{01}=3.2$), an equivalence test excludes
effects larger than $r\approx0.14$, and the target's split-half reliability ($\approx0.82$; ceiling
$r\approx0.9$) shows the null is not a noisy-label artifact. We release the code, the video-ID
manifest, and an acquisition method that works despite YouTube's SABR-only streaming.

\vspace{4pt}
\textbf{Date:}~\today \\
\textbf{Correspondence:}~\texttt{barada@gmail.com}, \texttt{cs21bt067.alum25@iitdh.ac.in}
}

\vspace{6pt}

\section{Introduction}
Encoding models that predict brain activity from naturalistic stimuli have improved sharply, with
deep multimodal architectures such as TRIBE winning the 2025 Algonauts challenge (out of 263 teams)
by mapping fused text, video, and audio features onto the cortical surface~\citep{dascoli2025tribe}.
Separately, the \emph{neuroforecasting} literature shows that \emph{measured} neural signals
(fMRI/EEG) can predict aggregate population behavior beyond self-report, from cultural
popularity~\citep{berns2012neural} to crowdfunding and market outcomes~\citep{genevsky2017brain},
and that the temporal reliability of neural processing tracks audience
preferences~\citep{dmochowski2014audience,hasson2004intersubject}.

Whether \emph{predicted} neural signals, which require no scanner and are inexpensive to compute,
inherit this predictive power has not been tested. The expected direction is not obvious. An
accurate encoder might preserve the behaviorally-relevant structure of the measured response; it
might equally regress that structure toward the group mean, discarding exactly the individual and
reward-region variability that the neuroforecasting effect depends on (\S\ref{sec:related}). We run
TRIBE without modification, reduce its prediction to a per-second engagement curve, and test that
curve against YouTube's ``most replayed'' heatmaps, and find no correlation. Establishing this meant
separating a genuine null from the confounds that would mimic one (temporal position, low-level
loudness and motion, the choice of readout), and then \emph{bounding} it with an equivalence test, a
Bayes factor, and a target reliability ceiling. One by-product is a cautionary result of its own: a
supervised cross-validated readout of the predicted cortex seems to predict re-watch at $r=0.47$,
yet this is entirely a shared temporal-shape artifact that vanishes under a stronger position
control, a warning for anyone using most-replayed as a label. It also meant obtaining the videos at all, which
YouTube's SABR streaming places beyond standard download tools. The code, the video-ID manifest, and
the acquisition pipeline are released.

The result is not only a null but an account of \emph{why} the transfer fails, which we take to be
the paper's main contribution. A group-trained encoder collapses idiosyncratic structure toward a
shared mean, and we see this same collapse from two directions: across videos, the weak
content-specific signal present in the visual inputs is absent from the predicted cortex
(\S\ref{sec:features}); and across subjects, per-subject predictions built to test the
inter-subject-reliability route, the closest prior positive result, carry no re-watch signal either
(\S\ref{sec:isc}). The averaging that makes these models accurate predictors of the \emph{typical}
response is what removes the variation a behavioral signal would ride on. Alongside this mechanism we
contribute a concrete methodological caution, the manufactured $r=0.47$ above, and a bounded rather
than merely unrejected null. We frame the work for practitioners tempted to use brain-encoding
models, or the foundation features behind them, as off-the-shelf engagement predictors.

\section{Related Work}\label{sec:related}
\paragraph{Brain encoding of naturalistic stimuli.} Voxel-wise and surface-based encoding models
map stimulus features onto fMRI responses~\citep{naselaris2011encoding,huth2016semantic}; the
Algonauts challenges have driven this to multimodal video~\citep{gifford2025algonauts}, and TRIBE
is the winning entry, fusing Llama-3.2~\citep{grattafiori2024llama},
V-JEPA 2~\citep{assran2025vjepa2}, and Wav2Vec-BERT and predicting per-TR responses on the
\texttt{fsaverage5} surface~\citep{dascoli2025tribe}.
\paragraph{Neuroforecasting.} Measured neural signals predict behavioral and market outcomes beyond
stated preference, but the relevant results differ in \emph{grain} from ours. Berns and
Moore~\citep{berns2012neural} and Genevsky et al.~\citep{genevsky2017brain} are \emph{cross-item}:
nucleus-accumbens and related reward activity, averaged over a stimulus, predicts which \emph{song}
or \emph{crowdfunding project} succeeds across the population, not which moment \emph{within} a clip
is re-watched. The closest analog for \emph{online content} is the valuation account of information
virality~\citep{scholz2017virality}, in which ventral-striatal and vmPFC/MPFC activity predicts which
New York Times articles are shared at scale; this too is cross-item, but it directly motivates the
signed vmPFC/MPFC readout we test (and find null). Only Dmochowski et
al.~\citep{dmochowski2014audience}, via inter-subject correlation of the neural time-course, speaks
to \emph{within-stimulus} moment-level engagement, and it depends on cross-subject reliability rather
than mean activation. Our target is within-video and moment-level, so the closest prior positive
result is the ISC line, not the reward-magnitude line.
A reason to expect transfer to \emph{predicted} signals is that an accurate encoder should reproduce
whatever behaviorally-relevant structure the response carries; two reasons to doubt it are that the
reward/salience signal is largely subcortical (Knutson's NAcc) and region-specific, which a
group-trained cortical encoder regresses toward the mean and a whole-cortex readout discards, and
that a model optimized for average fMRI accuracy need not preserve moment-to-moment contrasts. We
test which of these dominates.
\paragraph{Engagement prediction.} Video ``highlight'' and engagement models typically use
low-level audiovisual features; ``most replayed'' is a large but biased target (intro/onset
effects, chapter markers, seek-back behavior), which motivates our position and baseline controls.

\section{Method}
\subsection{Model}
TRIBE~\citep{dascoli2025tribe} is a 1B-parameter trimodal encoder trained on $500{+}$ hours of
fMRI from $700{+}$ individuals. It extracts features from three frozen foundation encoders
(Llama-3.2~\citep{grattafiori2024llama} over the dialogue transcript, V-JEPA 2~\citep{assran2025vjepa2}
over video frames, and Wav2Vec-BERT over the soundtrack), temporally aligns them, and fuses them
with a Transformer conditioned on a learned subject embedding to predict the per-TR cortical
response $\mathbf{P}\in\mathbb{R}^{T\times V}$ on the \texttt{fsaverage5} surface
($V{=}20{,}484$ vertices, TR${=}1$\,s). We use the released weights with no fine-tuning; because we
study relative temporal dynamics, we average over subject embeddings and analyze the resulting
predicted response.

\subsection{Engagement readout}
We summarize the high-dimensional prediction into a scalar per-TR \emph{engagement} value via the
\emph{global field power} (GFP), the root-mean-square over vertices,
\begin{equation}
e_t=\sqrt{\tfrac{1}{V}\textstyle\sum_{v=1}^{V} P_{t,v}^2},\qquad t=1,\dots,T .
\end{equation}
GFP indexes how strongly the stimulus drives the cortex overall and makes no assumption about
which regions matter. We take it as a candidate engagement signal and test that interpretation
rather than assume it; the region-restricted variants in \S\ref{sec:region} relax the
whole-cortex assumption. Each TR is placed on the true video timeline using the model's segment
onsets, and we retain the first $60$\,s ($\approx 60$ TRs) per video.

\subsection{Behavioral target}
YouTube's ``most replayed'' heatmap reports $100$ markers per video, each a normalized
re-watch intensity in $[0,1]$ (relative to the video's peak). We linearly interpolate the marker
series onto the model's TR grid to obtain a target $g_t$ commensurate with $e_t$.

\subsection{Correlation and position control}
The raw association is the Pearson correlation $r_{\text{raw}}=\mathrm{corr}(e,g)$. Because both the
engagement curve and most-replayed carry a strong low-order temporal trend (intros and onsets are
re-watched, and predicted response often decays over a clip), $r_{\text{raw}}$ conflates
\emph{content} with \emph{position}. We therefore use as the primary metric the
\emph{position-controlled partial correlation}: we regress each series on a position basis
$\mathbf{B}=[\mathbf{1},\,t,\,t^2]$ by ordinary least squares and correlate the residuals,
\begin{equation}
r_{\text{part}}=\mathrm{corr}\big(e-\mathbf{B}\hat{\beta}_e,\; g-\mathbf{B}\hat{\beta}_g\big).
\end{equation}
The quadratic basis removes the dominant monotone-plus-onset trend without overfitting $\sim\!60$
points, isolating whether $e$ tracks $g$ at the level of \emph{which specific moments} are
re-watched.

\subsection{Pooling and inference}
Each video yields one $r_{\text{part},i}$. We report the unweighted mean across videos (equivalent
to Fisher-$z$ pooling here, and avoids over-weighting longer clips), test it against zero with a
one-sample $t$-test and a sign test, and compare it to each baseline with a paired $t$-test. We
report 95\% confidence intervals from the between-video standard deviation.

\subsection{Low-level baselines}
To calibrate any observed effect, we compute two content-derived control curves and pass them
through the identical raw/partial pipeline: \emph{loudness}, the per-second RMS energy of the
mono $16$\,kHz waveform; and \emph{motion}, the mean absolute frame-to-frame pixel difference of
$1$\,fps, $64{\times}36$ grayscale frames. TRIBE is deemed predictive only if $r_{\text{part}}$
clearly exceeds these.

\section{System and Pipeline}
The study depends on two non-standard pieces of infrastructure: a means of obtaining the videos and
a means of encoding them at acceptable cost. YouTube's current streaming defeats common download
tools, and V-JEPA 2 encoding dominates runtime. We describe how the released pipeline addresses both.

\paragraph{Acquisition under SABR.} As of 2025 YouTube serves most popular videos via SABR
(server-side adaptive bitrate) streaming, which exposes no directly downloadable media; standard
tools (yt-dlp, youtube-dl, cobalt) return only storyboard images or fail, on both residential and
datacenter IPs. We instead acquire videos with the NewPipe Android client running on a physical
device, driven programmatically over ADB with UI-automation, which succeeds where those tools fail.
The behavioral target is unaffected: most-replayed heatmaps are metadata and are fetched separately
without downloading media.

\paragraph{Encoding cache.} Encoding is the dominant cost ($\sim\!6$--$13$ minutes of GPU per clip,
driven by V-JEPA 2). We cache the model output $\mathbf{P}$ (and per-TR onsets) on a network volume,
keyed by video and analysis window, so that no video is re-encoded across runs or re-analyses;
downstream readouts, baselines, and statistics are cheap CPU operations recomputed on demand.

\paragraph{Resumable, connectivity-independent scoring.} Scoring runs as a deployed serverless
function that fans out one video per GPU worker, each reading its clip from the volume; every
per-video result is committed immediately, so the study is resumable and survives client
disconnects, and a GPU-free aggregation step pools the cached results incrementally as they land.

\section{Experiments}
We analyze $N=48$ videos with most-replayed heatmaps spanning $11$ categories (music~17, talk~5,
tech~4, comedy~4, education~4, food~3, science~3, reaction~2, gaming~1, trailer~1, misc~4). For
each video we analyze the first $60$\,s ($\approx 60$ TRs) and compute the engagement curve, the
two low-level baselines, and the most-replayed target.

\section{Results}
\begin{figure}[h]
\centering
\includegraphics[width=\textwidth]{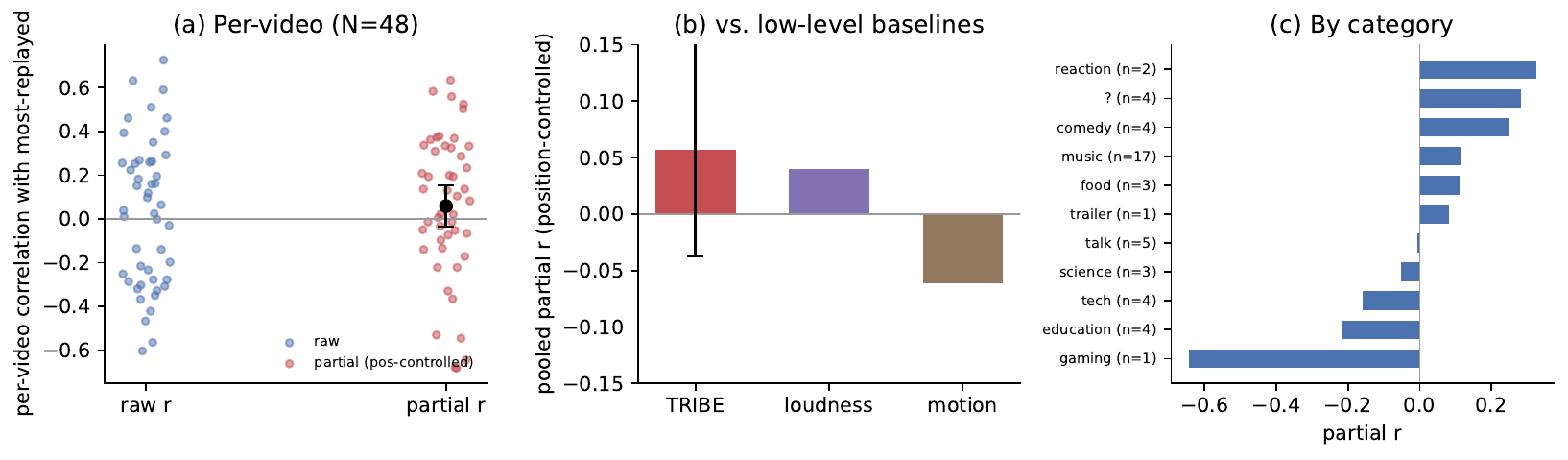}
\caption{\textbf{No content-level prediction of re-watch behavior.}
(a)~Per-video raw and position-controlled correlations with most-replayed; the partial correlation
(mean $\pm$ 95\% CI) is centered on zero and the CI crosses it.
(b)~Pooled partial correlation: TRIBE is statistically indistinguishable from the loudness baseline
and near zero.
(c)~Per-category partial correlations are small, sign-inconsistent, and dominated by noise at
small $n$.}
\label{fig:main}
\end{figure}

Table~\ref{tab:main} and Figure~\ref{fig:main} summarize the findings.

\begin{table}[h]
\centering
\begin{tabular}{lcc}
\toprule
signal & pooled raw $r$ & pooled partial $r$ (position-controlled) \\
\midrule
\textbf{TRIBE engagement} & $+0.036$ & $\mathbf{+0.058}$ \\
loudness baseline & --- & $+0.040$ \\
motion baseline & --- & $-0.061$ \\
\bottomrule
\end{tabular}
\caption{Pooled correlations of TRIBE engagement and low-level baselines with YouTube
most-replayed ($N=48$). The TRIBE partial correlation is not significantly different from zero
($t(47)=1.21$, $p=0.23$; 95\% CI $[-0.04,0.15]$) and not significantly greater than the loudness
baseline (paired $p=0.74$).}
\label{tab:main}
\end{table}

\paragraph{Primary test.} The pooled TRIBE position-controlled partial correlation is $+0.058$
(between-video SD $=0.33$; 95\% CI $[-0.04,0.15]$), not significantly different from zero
(one-sample $t(47)=1.21$, $p=0.23$; sign test $28/48$ positive, $p=0.25$).
\paragraph{Baseline comparison.} TRIBE does not exceed the low-level baselines: the paired
difference TRIBE${}-{}$loudness ${}=0.018$ ($t=0.34$, $p=0.74$); the motion baseline is $-0.06$.
\paragraph{Raw correlation is $\approx 0$ on diverse content.} The pooled raw correlation is also
$\approx 0$ ($+0.036$). The moderate raw correlations ($0.3$--$0.8$) seen for music videos are a
genre-specific intro/onset-replay artifact that vanishes on talks, tech, science, comedy, etc.
\paragraph{Per-category.} Category-level partial correlations are small and inconsistent
(comedy $+0.25$, music $+0.11$, education $-0.21$, science $-0.05$; several $n{=}1$); no content
type shows systematic prediction.

\subsection{Video-level ranking}
The analysis above asks whether TRIBE predicts \emph{which moments} within a video are re-watched.
A complementary question is whether TRIBE ranks \emph{which videos} are more engaging overall. We
correlate video-level TRIBE summaries (mean and peak engagement) with public engagement metrics
(view and like counts) across the same $48$ videos. All associations are near zero and, if
anything, slightly negative: Spearman $\rho(\text{mean},\text{views})=-0.09$,
$\rho(\text{mean},\text{likes})=-0.14$, $\rho(\text{peak},\text{views})=-0.20$, and
$\rho(\text{mean},\text{like/view})=-0.08$ (all $|\rho|<0.28$, the $p{=}0.05$ threshold at
$n{=}48$; none significant). TRIBE engagement is thus not an indicator of video-level engagement
either. We note an important limitation: our videos are all already highly popular (views
$8\!\times\!10^4$ to $9\!\times\!10^9$, median $1.3\!\times\!10^7$), because they require a
most-replayed heatmap; this range restriction weakens the test, and a definitive virality study
would need a balanced viral-vs-flop sample.

\subsection{Region-specific readouts}\label{sec:region}
The global-field-power readout pools over the whole cortex and could dilute a signal carried by a
specific functional network, plausibly the salience/reward or sensory systems implicated in
neuroforecasting. We therefore repeated the position-controlled analysis with the engagement curve
restricted to each of five networks defined by the Destrieux atlas, recomputing GFP over only the
vertices of that network. The null is robust across all of them (pooled partial $r$, $n{\approx}48$):
whole-cortex $+0.058$, visual $-0.010$, auditory $+0.065$, salience (insula/cingulate) $+0.001$,
frontal $+0.023$, and parietal $+0.088$. The largest value (parietal) is marginal and would not
survive correction for the six readouts tested; no network approaches a level that would overturn
the whole-cortex conclusion. Because the GFP is an unsigned magnitude while the neuroforecasting
literature concerns \emph{signed} activation in \emph{specific} value/salience regions, we also
computed the \emph{signed mean} response in canonical surface ROIs (vmPFC/MPFC, anterior cingulate,
anterior insula). These are null as well: vmPFC $+0.088$, ACC $+0.015$, anterior insula $+0.037$
(all $95\%$ CIs cross zero, $n{=}48$). Spatially resolving the predicted response, signed or
unsigned, does not recover a content-level re-watch signal. One caveat here is anatomical: the
nucleus accumbens / ventral striatum, a primary neuroforecasting locus, is
subcortical and \emph{absent} from the \texttt{fsaverage5} surface mesh, so this test reaches at
most its cortical shadow (\S\ref{sec:limits}).

\subsection{Robustness: temporal permutation}
Because both series are autocorrelated, a standard parametric test can be anti-conservative. As a
non-parametric check we recomputed the pooled partial correlation under a circular-shift null that
preserves each engagement curve's autocorrelation structure while destroying its temporal alignment
to most-replayed ($K{=}2000$ shifts, $n{=}48$ videos). The observed pooled partial $r=0.058$ falls
well within the null distribution (two-tailed $p=0.12$), consistent with the parametric test and
confirming that the near-zero effect is not an artifact of temporal autocorrelation.

\subsection{Strength of evidence for the null}\label{sec:equiv}
A non-significant $p$ is not evidence of absence, so we quantify the evidence directly. A
two-one-sided-tests (TOST) procedure against a pre-specified smallest effect of interest
$\delta=0.10$ does \emph{not} reach equivalence ($p=0.20$): we cannot rule out a true effect smaller
than $r\approx0.1$, the limit of resolution at $n{=}48$. Descriptively, the $90\%$ CI is
$[-0.02,0.14]$, so effects larger than $r\approx0.14$ are excluded. A default JZS Bayes factor for
the one-sample test gives
$\mathrm{BF}_{01}=3.2$: the data are about three times more likely under the null than under a
diffuse alternative, which is positive evidence \emph{for} no effect. A random-effects
meta-analysis (DerSimonian--Laird over per-video Fisher-$z$) leaves the estimate unchanged
($r=0.058$) and, because between-video heterogeneity is high ($I^2=87\%$), does not narrow the
interval; this locates the limiting factor squarely at the sample size rather than the pooling
method, and motivates the larger, balanced sample we flag as the natural next step.
This is not a low-ceiling artifact. Estimating the reliability of the most-replayed target by
split-half correlation of its $100$ markers (odd vs.\ even, Spearman--Brown corrected) gives a
median reliability of $0.82$, so even a perfect predictor is bounded near $r\approx0.9$; the target
is highly reliable for typical clips and the near-zero correlation reflects the predictor, not
target noise. Reliability is heterogeneous, though: the $29\%$ of clips whose $60$\,s window
contains fewer than $10$ native markers have coarser targets (\S\ref{sec:confound}).

\subsection{A learned readout does not rescue the null}\label{sec:probe}
The GFP is a fixed, unsupervised scalar; perhaps a \emph{supervised} readout of the predicted
cortex could extract a signal it discards. We test this with a cross-validated probe: the
$20{,}484$-vertex predicted response is reduced to $100$ principal components (fit within each
training fold), ridge regression is fit to per-TR most-replayed, and performance is evaluated under
grouped \textbf{leave-one-video-out} cross-validation (group $=$ video, so no within-video
leakage), pooling per-video correlations as before. Under a quadratic position control this probe
appears to succeed, reaching pooled CV $r=0.47$. This figure is not comparable to the
GFP's per-video partial $r=0.058$ and does not imply a signal the GFP missed: a supervised,
cross-validated readout can learn a temporal shape common to the training videos and apply it to a
held-out one, which a single fixed per-video curve cannot, so a much larger number is expected even
under the null. And indeed it does not survive scrutiny. A
control that ignores the cortex entirely and predicts only the \emph{average} residual
most-replayed curve of the training videos reaches $r=0.47$ as well, and predicting a held-out
video's curve from a \emph{different} video's probe output (its prediction paired with every other
video's target, averaged) still yields $r=0.32$: the probe is
reproducing a temporal \emph{shape} shared across videos, not video-specific content. When the
position control is strengthened from a quadratic to a cubic-spline basis ($7$ df), the probe
collapses to $r=0.14$ and its predictions are no more correlated with a video's own most-replayed
curve than with an unrelated video's (matched $0.14$ vs.\ mismatched $0.15$; paired $p=0.87$). A
supervised learned readout of the predicted cortex therefore carries \emph{no} content-specific
re-watch signal; the apparent success is the intro/onset-shape artifact operating at higher
temporal order. This is a concrete caution for using most-replayed as a training label: a coarse
detrend manufactures a large cross-validated correlation that is entirely spurious.

\subsection{Where the signal is lost: input features vs.\ predicted cortex}\label{sec:features}
The cortical probe's failure poses a mechanistic question: is a content-specific signal simply
absent from the stimulus, or present in the model's inputs but discarded by the fMRI-encoding step?
We run the identical probe, same spline position control and matched/mismatched test, on each of
TRIBE's three input streams, captured in-model at the modality projectors: V-JEPA 2 video,
Wav2Vec-BERT audio, and Llama text. Only the \textbf{visual} stream shows any video-specificity, and
only weakly: matched-target correlation $0.12$ versus mismatched $0.05$
(paired $p=0.06$). The audio (matched $-0.01$ vs.\ mismatched $0.02$, $p=0.43$) and text (matched
$0.00$ vs.\ mismatched $0.02$, $p=0.74$) streams show none, behaving like the predicted cortex
(matched $0.14 \approx$ mismatched $0.15$, $p=0.87$). Whatever content-specific signal exists lives
in the visual input alone; audio and text carry nothing for the encoding step to lose.

This visual signal is small and borderline, and we are deliberately cautious about it. With a
separately extracted V-JEPA 2 representation (a different checkpoint and denser temporal sampling)
the same visual probe reaches matched $0.13$ vs.\ mismatched $0.00$ ($p=0.004$); across the two
extractions the effect sits at $p\approx0.004$--$0.06$. We therefore treat it as suggestive at best:
a hint of video-specific structure in the visual features that the predicted cortex does not carry. Even so, the comparison must be read off the \emph{mismatched} baselines, not
the matched ones. The cortex probe's matched correlation ($0.14$) is in fact slightly \emph{higher}
than the visual probe's ($0.12$); what differs is mismatched, where the cortex ($0.15$) matches any
video equally while the visual features ($0.05$) fit mainly the correct one. Matched and mismatched
run through an identical pipeline, differing only in the pairing: each held-out video's prediction is
resampled to a common $50$-point normalized-time grid and correlated with its \emph{own} target
(matched) or, averaged, with every \emph{other} video's target on the same grid (mismatched), so the
gap cannot be a grid or normalization artifact. To the extent the effect is real, it is the ``regress
toward the mean'' tendency of \S\ref{sec:related}: a group-trained encoder collapses
idiosyncratic responses toward a shared mean, so the predicted cortex keeps the shape but not the
faint video-specific structure the visual features carry. The finding is modest by construction: the
signal is weak (pooled CV $r\approx0.12$, far below the target's $r\approx0.9$ ceiling) and confined
to one of three streams, so it locates \emph{where} a small signal would be lost rather than
recovering a strong one.

\subsection{Testing the inter-subject-correlation route directly}\label{sec:isc}
The one prior positive result at our grain, inter-subject correlation (ISC) of the neural
time-course (\S\ref{sec:related}), requires \emph{per-subject} responses. The released
\texttt{facebook/tribev2} checkpoint cannot supply them: it was trained with subject averaging
(\texttt{average\_subjects}\,=\,True) and contains no subject-specific parameters (no
subject-embedding layer, subject-layer \texttt{n\_subjects}\,$=$\,$0$, empty subject mapping), so its
predictions for the four Algonauts subjects are bit-identical. Rather than concede the route, we
build the per-subject responses ourselves and test ISC end-to-end.

\paragraph{Per-subject encoders.} Using the public Algonauts~2025 / CNeuroMod fMRI (four subjects
watching Friends, cortical Schaefer-1000 parcels, TR${=}1.49$\,s), we fit a ridge encoder per subject
that maps foundation features (V-JEPA~2, optionally with Wav2Vec-BERT audio) plus hemodynamic delays
onto the $1{,}000$ parcels. Under leave-one-episode-out cross-validation these encoders predict
held-out fMRI at a mean parcel correlation of $r\approx0.15$, in the published range for
feature-to-cortex encoders. They also \emph{generalize across video domains}: an encoder trained only
on Friends (situation comedy) predicts held-out fMRI for feature films (\emph{Bourne}, \emph{Wolf of
Wall Street}) at $r\approx0.10$, nearly the in-domain level, which is direct evidence that the
encoders transfer to unseen naturalistic video rather than memorizing one genre.

\paragraph{Predicted ISC does not track re-watch.} We apply the four per-subject encoders to each
re-watch video's features, obtaining four predicted cortical responses per video; the per-TR
predicted ISC is the mean pairwise across-subject correlation of the predicted parcel patterns. Under
the same position-controlled protocol, this predicted-ISC curve does not predict most-replayed:
pooled partial $r=-0.04$ (video features; $p=0.34$, $n=48$) and $r=-0.03$ (video${+}$audio;
$p=0.48$). The null holds across both feature sets and all four subjects, with encoders that are
validated in-domain and cross-domain. Even the closest prior positive result to our target, computed
with genuine per-subject predictions rather than the subject-averaged model, provides no
content-level prediction of what viewers re-watch. (We omit TRIBE's text stream from these encoders:
it is the null modality of \S\ref{sec:features} and, via word-level language-model features, by far
the most expensive to extract; a full three-stream encoder is left to future work but is not expected
to change a null that already holds for the informative streams.)

\subsection{No duration or resolution confound}\label{sec:confound}
Because most-replayed's $100$ markers span the full clip, a short video contributes many in-window
markers and a long one few; if apparent signal tracked marker density it would be a resolution
artifact entangled with category. It does not: across videos the raw correlation is uncorrelated
with in-window marker density (Spearman $\rho=-0.08$) and with clip duration ($\rho=0.07$), and the
partial correlation is flat in density ($\rho=0.12$). The median clip has $20$ in-window native
markers; the $29\%$ with fewer than $10$ have coarser targets but do not drive the result.

\begin{figure}[t]
\centering
\includegraphics[width=\textwidth]{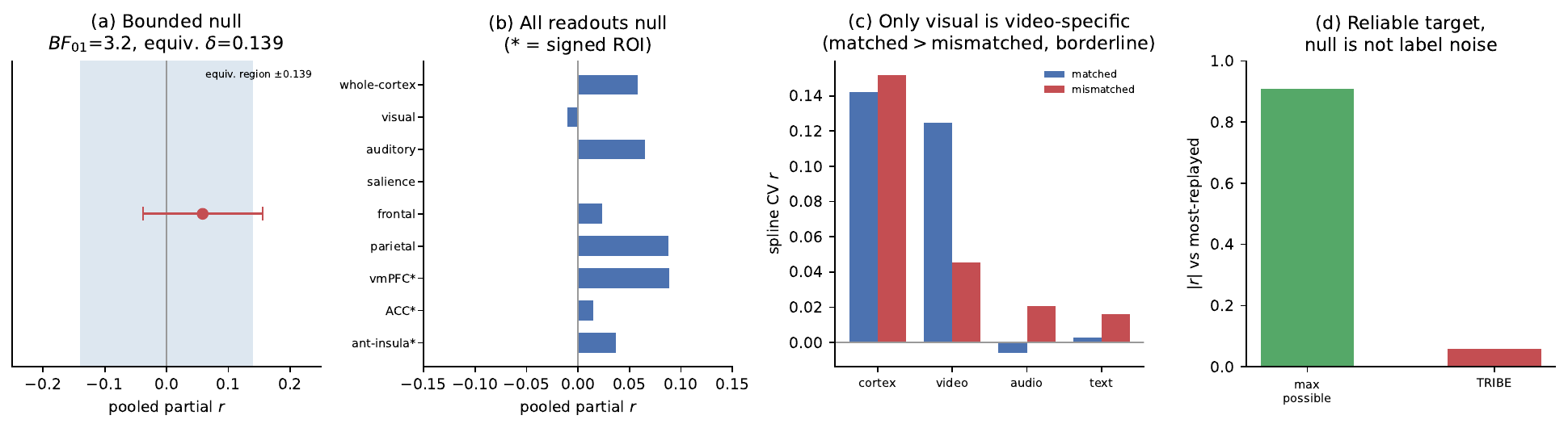}
\caption{\textbf{Bounding the null, and locating where the signal is lost.}
(a)~The pooled partial correlation (point $\pm$ 95\% CI) lies inside an equivalence region of
$\delta{=}0.14$; a Bayes factor gives $\mathrm{BF}_{01}{=}3.2$ (moderate evidence for the null).
(b)~Every readout is near zero: whole-cortex GFP, five functional networks, and signed
value/salience ROIs ($*$).
(c)~Under the spline position control, matched vs.\ mismatched CV correlation by source: only
TRIBE's visual input stream shows any video-specificity (matched $>$ mismatched, borderline
$p{=}0.06$); the audio and text streams and the predicted cortex show none (matched $\approx$
mismatched). The apparent $r{=}0.47$ of the cortex probe under a coarse quadratic control (not shown)
is a shared-shape artifact that collapses here.
(d)~The target is reliable (split-half ceiling $r\approx0.9$), so the near-zero result is not label
noise.}
\label{fig:evidence}
\end{figure}

\section{Discussion}
The apparent signal is fragile at every level we probed. In a music-only pilot the raw correlation
was moderate to strong ($0.3$--$0.8$), but it survives neither the extension to non-music content
nor a first-order position control; and when a supervised readout appears to recover a large effect
($r=0.47$), that too dissolves into a shared temporal-shape artifact under a stricter position
control and a matched-versus-mismatched test. Methodologically, most-replayed
carries a dominant, genre- and position-linked temporal shape, and any analysis that does not
aggressively remove it (raw correlation, a coarse detrend, or a cross-validated probe on top of a
coarse detrend) will report a spurious signal. This is a direct caution for the growing practice of
repurposing brain-encoding models as off-the-shelf engagement predictors. The faint dissociation
between the visual input and the predicted cortex points the same way. What little content-specific
signal appears anywhere (a borderline effect in the visual stream, nothing in audio, text, or the
predicted cortex) is not carried by the encoded response, which places the loss at the
fMRI-encoding step and not the readout or the target. Both observations trace to one property of a
group-trained encoder: it collapses idiosyncratic structure toward a shared mean. That happens
across videos, so the weak video-specific structure in the visual features survives in
neither the predicted cortex nor the shared temporal shape the probe latches onto
(\S\ref{sec:features}); and across subjects, so completely that the released checkpoint carries no
per-subject parameters and cannot express inter-subject differences at all (\S\ref{sec:isc}). The
averaging that makes these models accurate for the \emph{typical} cortical response is also what
strips out the individual and moment-specific variation a behavioral signal would need.

The surrounding evidence is what gives the null its weight. The target is
reliable (split-half $\approx0.82$; ceiling $r\approx0.9$), so the near-zero result is not a
noisy-label artifact; a Bayes factor gives moderate evidence \emph{for} the null
($\mathrm{BF}_{01}=3.2$); an equivalence test rules out effects larger than $r\approx0.14$; and the
result is stable under an autocorrelation-preserving permutation null. We can therefore state a
narrow but definite claim. On this target and these readouts, predicted-fMRI drive from TRIBE
carries \emph{approximately no} content-specific re-watch signal, up to a bound of $r\approx0.14$;
we do not claim the effect is exactly zero.

\section{Limitations}\label{sec:limits}
Several limitations bound the scope, and two matter most. First, \texttt{fsaverage5} is a
\emph{cortical-surface} mesh: the nucleus accumbens / ventral striatum, a primary neuroforecasting
locus, is subcortical and simply absent, so TRIBE-on-\texttt{fsaverage5} structurally cannot
represent the canonical ventral-striatal reward signal and we test at most its cortical shadow. The
natural follow-up is a subcortical-ROI encoder fit on the volumetric CNeuroMod BOLD (which includes
the nucleus accumbens); that data is under registered access, so we leave it as immediate future
work with the encoder pipeline in place. Second, most-replayed is a biased behavioral target (onset/intro
effects, chapter markers, seek-back), and although we control position aggressively and confirm the
target is reliable, a cleaner signal such as creator-side audience-retention would be preferable.
Further: we analyze a $60$\,s window; our videos are all already popular (they require a heatmap),
a range restriction that weakens the video-level ranking test; our ISC test (\S\ref{sec:isc}) uses
per-subject encoders we fit rather than TRIBE's own per-subject head (which the released model lacks),
and covers the four Algonauts subjects; and $N=48$ with an imbalanced category mix is modest, though
our pipeline makes scaling to hundreds of videos cheap.

\section*{Ethics and Broader Impact}
A reliable predictor of moment-level engagement from brain-derived signals would have clear
dual-use potential for attention-optimized and manipulative media. Our finding is negative, which
reduces that concern for this specific approach, but the acquisition and scoring pipeline we release
is general; we release it for scientific reproducibility and note that engagement-optimization
applications warrant care. No human-subjects data were collected; most-replayed heatmaps are public
aggregate metadata and no per-user data is used.

\section{Conclusion}
For this target and these readouts, a predicted-fMRI drive signal from TRIBE does not forecast
YouTube re-watch behavior beyond temporal position and low-level features, and the null is backed
by an equivalence bound, a Bayes factor, a target reliability ceiling, and the failure of a
supervised readout that a naive detrend made look successful. The story is not quite ``no signal
anywhere''. A small, borderline content-specific signal appears in the visual input stream, though
not in audio, text, or the predicted cortex, which tentatively places what little exists upstream of
the brain-mapping stage. We keep the scope narrow: a
single model, a cortical-surface readout that excludes subcortical reward regions, a biased target,
and a $60$\,s window. Whether a subcortex-inclusive encoder, a per-subject reliability readout, or a
cleaner behavioral target would change the result remains open; code and video IDs are released to
enable those tests.

\section*{Reproducibility}
Code (scoring, position-controlled validation, baselines, SABR-resilient acquisition, encoding
cache), the manifest of YouTube video IDs, and per-video results are available at
\url{https://github.com/mercurialsolo/tribe-replay-heatmaps}. We do not redistribute video or fMRI
data; the most-replayed heatmaps are public YouTube metadata fetched per ID, so the full analysis
is reproducible from the released IDs and code.

\end{document}